\documentclass[12pt]{iopart}

\usepackage{amssymb}
\usepackage{epsfig}
\usepackage{setstack}

\begin{document}

\title{Exact cosmological solutions of scale-invariant gravity theories}

\author{John D Barrow and T Clifton}

\address{DAMTP, Centre for Mathematical Sciences, University of
  Cambridge,  Wilberforce Road, Cambridge, CB3 0WA, UK.}

\eads{\mailto{J.D.Barrow@damtp.cam.ac.uk},
\mailto{T.Clifton@datmp.cam.ac.uk}}

\pacs{98.80.Jk, 04.20.Jb}

\begin{abstract}
We have found new anisotropic vacuum solutions for the scale-invariant
gravity theories which generalise Einstein's general relativity to a theory
derived from the Lagrangian $R^{1+\delta }$. These solutions are expanding
universes of Kasner form with an initial spacetime singularity at $t=0
$ and exist for $-1/2<\delta <1/4$ but possess different Kasner index
relations to the classic Kasner solution of general relativity to which they
reduce when $\delta =0$. These solutions are unperturbed by the introduction
of non-comoving perfect-fluid matter motions if $p<\rho $ on approach to the
singularity and should not exhibit an infinite sequence of chaotic Mixmaster
oscillations when $\delta >0$.
\end{abstract}

\maketitle

There have been many studies of gravity which consider generalisations of
the Einstein-Hilbert action of general relativity by the addition of terms
of higher-order in the curvature to the Lagrangian, see refs in \cite{BO,
schmidt}. These are of interest for the evolution of cosmological models
close to the initial singularity and at late times, for the study of quantum
gravitational phenomena, inflation, and the observational consequences on
solar system scales. In particular, it is possible to study the structure of
gravity theories in which the Lagrangian is quadratic in the curvature or an
arbitrary analytic function of the curvature \cite{kerner, BO, ruz}.
Unfortunately, such studies are complexified by the lack of exact solutions
of the field equations for these theories other than the special cases
provided by the solutions of general relativity. The field equations are
typically fourth order and difficult to deal with except for spacetimes with
isotropy and homogeneity or static spherical symmetry. We have recently
considered in detail a different type of generalisation of Einstein's theory
which derives from a power-law Lagrangian of the form $R^{1+\delta}$, where $%
\delta $ is constant. This theory reduces to general relativity when $\delta
\rightarrow 0$, and both the flat Friedmann universes \cite{schmidt,
  Clif, Car04} and static spherically symmetric solutions \cite{Clif} can be found,
tested for stability, and confronted with cosmological and
perihelion-precession observations to produce tight observational limits on $%
\delta $ of \cite{Clif}

\[
0\leq \delta < 7.2\times 10^{-19}. 
\]

In this letter we will show that it is also possible to find exact
anisotropic and spatially inhomogeneous cosmological solutions of this
theory of Kasner type. These seem to be the first exact anisotropic
solutions of higher-order Lagrangian theories of gravity and provide a
testing ground for a range of interesting investigations. Kasner solutions 
\cite{kasner, LK} form the building blocks for an understanding of the
behaviour of the most general known solutions of the Einstein equations and
allow us to understand the conditions under which chaotic behaviour arises
and persists on approach to the initial cosmological singularity \cite{mis,
bkl, jb}. They also provide a simple environment in which to study quantum
gravitational effects like particle production \cite{zs}, and the conditions
under which shear anisotropy dominates the expansion of the universe at
early and late times. The simplicity of these new solutions is appealing and
allows for a complete understanding of their behaviour as $\delta $ varies.
They also provide an instructive context in which to evaluate the nature of
the general relativistic evolution and to determine whether it is typical or
atypical within this wide class of theories.

We consider here a gravitational theory derived from the Lagrangian density 
\begin{equation}
\mathcal{L}_{G}=\frac{1}{\chi }\sqrt{-g}R^{1+\delta },  \label{action}
\end{equation}%
where $\delta $ and $\chi $ are constants. The limit $\delta \rightarrow 0$
gives us the familiar Einstein--Hilbert Lagrangian of general relativity and
we are interested in the observational consequences of $\left\vert \delta
\right\vert $ $>0$. There is a conformal equivalence between this theory and
general relativity with a scalar field possessing an exponential
self-interaction potential \cite{CB, maeda}.

We denote the matter action as $S_{m}$ and ignore the boundary term.
Extremizing 
\[
S=\int \mathcal{L}_{G}d^{4}x+S_{m}, 
\]%
with respect to the metric $g_{ab}$ then gives \cite{Buc70} 
\begin{eqnarray}
\fl \delta (1-\delta ^{2})R^{\delta }\frac{R_{,a}R_{,b}}{R^{2}}-\delta
(1+\delta )R^{\delta }\frac{R_{;ab}}{R}+(1+\delta )R^{\delta }R_{ab} \\
-\frac{1}{2}g_{ab}RR^{\delta } -g_{ab}\delta (1-\delta ^{2})R^{\delta }\frac{%
R_{,c}R_{,}^{\ c}}{R^{2}}+\delta (1+\delta )g_{ab}R^{\delta }\frac{\Box R}{R}%
=\frac{\chi }{2}T_{ab},  \label{field}
\end{eqnarray}
where $T_{ab}$ is the energy--momentum tensor of the matter, and is defined
in the usual way. We take the quantity $%
R^{\delta }$ to be the positive real root of $R$ throughout.

We seek solutions of these equations for the Bianchi type I metric

\begin{equation}
ds^{2}=-dt^{2}+t^{2p_{1}}dx^{2}+t^{2p_{2}}dy^{2}+t^{2p_{3}}dz^{2}
\label{met}
\end{equation}%
where the Kasner indices, $p_{i},$ are constants. This metric is a solution
of the vacuum field equations of the $R^{1+\delta }$ theory if and only if
the Kasner indices satisfy the algebraic relations

\begin{eqnarray}
(1-\delta)(p_{1}^{2}+p_{2}^{2}+p_{3}^{2})+3\delta (1+2\delta ) =(1+2\delta
^{2})(p_{1}+p_{2}+p_{3}),  \label{con1} \\
(1-\delta )(p_{1}p_{2}+p_{1}p_{3}+p_{2}p_{3})+3\delta ^{2}(1+2\delta )
=\delta (2+\delta )(p_{1}+p_{2}+p_{3})  \label{con2}
\end{eqnarray}%
These constraints can be solved to yield two classes of solutions. The first
has

\begin{eqnarray*}
\sum_{i=1}^{3}p_{i}&=&\frac{3\delta (1+2\delta )}{(1-\delta )} \\
\sum_{i=1}^{3}p_{i}^{2} &=&\frac{3\delta ^{2}(1+2\delta )^{2}}{(1-\delta
)^{2}}.
\end{eqnarray*}%
These are only solved by the isotropic solution with

\[
p_{1}=p_{2}=p_{3}=\delta \frac{(1+2\delta )}{(1-\delta )}. 
\]%
This is the zero-curvature vacuum Friedmann universe found by Bleyer and
Schmidt \cite{schmidt1, schmidt}. The second class of solutions to (\ref%
{con1})-(\ref{con2}) is new and generalises the anisotropic Kasner universe
of general relativity to

\begin{eqnarray}
\sum_{i=1}^{3}p_{i} &=&1+2\delta  \label{con3} \\
\sum_{i=1}^{3}p_{i}^{2} &=&1-4\delta ^{2}.  \label{con4}
\end{eqnarray}%
Note that when $\delta =0$ this reduces to the standard Kasner solution of
general relativity \cite{kasner}. The first constraint (\ref{con3}) means
that we must have $\delta >-1/2$ for an expanding universe. The second
constraint (\ref{con4}) requires $-1/2<\delta <1/2$ for consistency. When $%
\delta $ falls outside this range then anisotropic solutions with this
simple power-law form do not exist. These constraints restrict the range of
variation of the individual $p_{i}$ as follows (assuming with out loss of
generality that they are ordered $p_{1}<p_{2}<p_{3}$):

\begin{equation}
\frac{1+2\delta -2\sqrt{(1+2\delta )(1-4\delta )}}{3}\leq p_{1}\leq \frac{%
1+2\delta -\sqrt{(1+2\delta )(1-4\delta )}}{3}  \label{p1}
\end{equation}

\begin{equation}
\frac{1+2\delta -\sqrt{(1+2\delta )(1-4\delta )}}{3}\leq p_{2}\leq \frac{%
1+2\delta +\sqrt{(1+2\delta )(1-4\delta )}}{3}  \label{p2}
\end{equation}

\begin{equation}
\frac{1+2\delta +\sqrt{(1+2\delta )(1-4\delta )}}{3}\leq p_{3}\leq \frac{%
1+2\delta +2\sqrt{(1+2\delta )(1-4\delta )}}{3}.  \label{p3}
\end{equation}%
These ranges are plotted in Figure 1 for different values of $\delta .$ They
impose the further constraint $-1/2<\delta <1/4$ on the range of allowed
values of $\delta $ for which real-valued solutions exist. A horizontal line
of constant $\delta $ intersects the two closed curves at four points which
give three line segments covering the allowed ranges of the Kasner indices
for all $-1/2\leq \delta \leq 1/4$. For $\delta =0$, the intersects give the
ranges for the Kasner solution in general relativity: $-1/3\leq p_{1}\leq
0\leq p_{2}\leq 2/3\leq 1$.

\begin{figure}[tbp]
\centerline{\epsfig{file=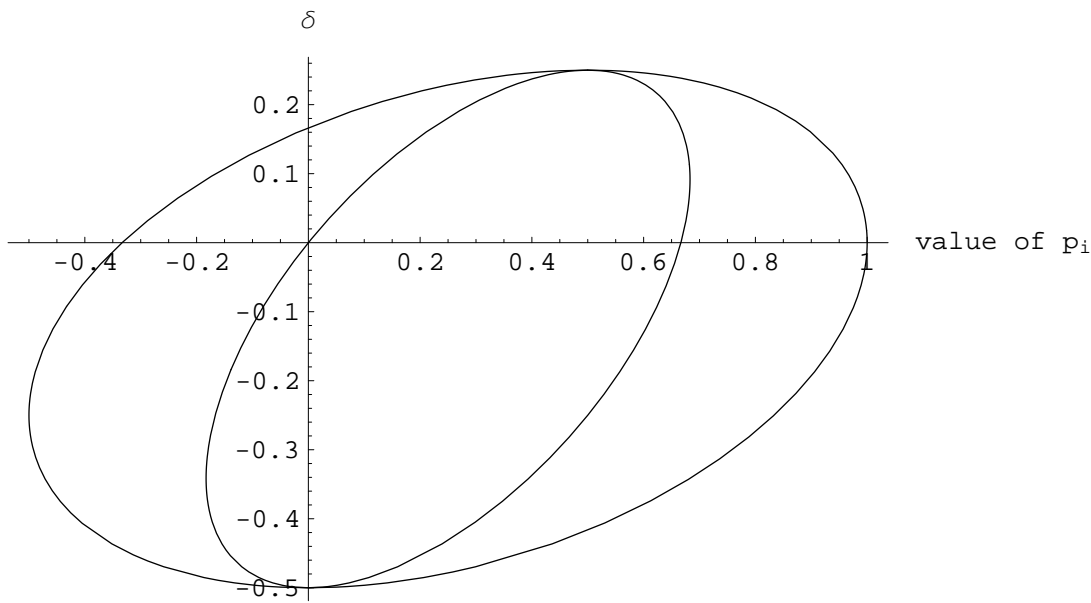,height=9cm}}
\caption{\textit{The intervals in which the Kasner indices $p_{i}$ lie can
be read off this graph. For any value of $\protect\delta $ in the range $-%
\frac{1}{2}<\protect\delta <\frac{1}{4}$ a horizontal line is drawn; the
boundaries of the intervals in which the $p_{i}$ lie are then given by the
four points at which the horizontal line crosses the two closed curves
defined by (\ref{p1})-(\ref{p3}). For $%
\protect\delta =0$ these boundaries can be seen to be $-\frac{1}{3}$, $0$, $%
\frac{2}{3}$ and $1$, as expected for the Kasner solution in general
relativity.}}
\end{figure}

We note the following features of these solutions, they have a
Weyl curvature singularity at $t=0$ so long as $\delta >-1/2$. Solutions
with $\delta <-1/2$ are contracting universes and have a `big rip'
singularity as $t\rightarrow \infty $. When $\delta >0$ it is possible for
all the Kasner indices to take positive values for every positive value of $%
\delta $.\ When $\delta <0$ we see that $p_{1}<0$ always and $p_{2}$ can
also take negative values but the universe expands in volume. The measure of
the positive values for $p_{1}$ increases as $\delta $ increases to its
maximum allowed value. This situation is of particular interest in
connection with the behaviour of more general Bianchi type IX Mixmaster
cosmologies in these gravity theories. In general relativity we know that
these cosmologies exhibit chaotic oscillatory behaviour on approach to the
initial Weyl curvature singularity at $t=0$ in \ vacuum \cite{mis, bkl,
jbprl, jb}. In order for an infinite number of chaotic oscillations to occur
we need one of the Kasner indices to take only negative values. This is the
case in vacuum in general relativity. When a scalar field is added the
Kasner relations change and it is possible for all indices to be positive
and the oscillatory sequence of permutations of the scale factors will cease
after a finite number of oscillations \cite{bkl2}. The subsequent evolution
will be in general of Kasner type. We expect a similar sequence of events in
the Mixmaster cosmologies in $R^{1+\delta }$ gravity theories when $\delta
>0 $. Oscillations of the scale factors as $t\rightarrow 0$ will eventually
permute the Kasner indices into one of the combinations in which they are
all positive and oscillations will cease leaving the evolution to proceed in
Kasner form. By contrast, we expect chaotic oscillations to persist when $%
-1/2<\delta <0$ although their detailed structure \cite{Che83} will differ from those
that occur in general relativity and will be investigated elsewhere.

The new form of the Kasner index constraints found here differs from the
form found in general relativity when a massless scalar field is added \cite%
{jb2, jb3} because a change is made to the sum of the indices (\ref{con3})
as well as to the sum of their squares, (\ref{con4}).

We can also investigate the extent to which this vacuum metric (\ref{met})
is a stable asymptotic solution of the field equations in the presence of
perfect-fluid matter as $t\rightarrow 0$ as it is in general relativity. We
need to determine whether the form of the exact solution leads to fluid
stresses which grow faster than the vacuum anisotropy terms (which are all $%
O(t^{-2(1+\delta )})$ as $t\rightarrow 0$). Similar investigations can be
made \textit{mutatis mutandis }for the case of fluids with anisotropic
pressures \cite{skew}.

Consider fluid motions in the background Kasner metric and assume that the
material content of the universe is a perfect fluid with equation of state $%
p=(\gamma -1)\rho $ with $1\leq \gamma <2$. The continuity equation is \cite%
{LL, jb2, jb4}

\begin{equation}
\frac{\partial }{\partial x^{i}}(t^{p_{1}+p_{2}+p_{3}}\rho ^{1/\gamma
}u^{i})=0,  \label{mo1}
\end{equation}%
where $u^{i}$ is the normalised 4-velocity ($u_{a}u^{a}=1$), and the
momentum conservation equation is

\begin{equation}
\ (\rho +p)u^{k}(u_{i},_{k}-\frac{1}{2}u^{l}g_{kl},_{i})=-\frac{1}{3}\rho
_{,i}-u_{i}u^{k}p_{,k}.  \label{mo2}
\end{equation}%
Neglecting the space derivatives with respect to the time derivatives, so
that we confine attention to scales larger than the particle horizon in the
velocity-dominated approximation, we have

\begin{equation*}
t^{p_{1}+p_{2}+p_{3}}u_{0}\rho ^{1/\gamma } =const \; ; \qquad u_{\alpha }\rho ^{(\gamma -1)/\gamma } =const.
\end{equation*}%
For relativistic motions, keeping the dominant velocity component ($%
u^{3}=u_{3}t^{-2p_{3}}$ and $u_{0}^{2}\sim u_{3}u^{3}\sim
(u_{3})^{2}t^{-2p_{3}}$) as $t\rightarrow 0$ gives to leading order $\rho
\sim t^{-\gamma (p_{1}+p_{2})/(2-\gamma )}$ and $u_{\alpha }\sim
t^{(p_{1}+p_{2})(\gamma -1)/(2-\gamma )}$. Using these asymptotic forms we
can now check that all the components of $T_{a}^{b}$ which they induce on
the right-hand side of the field equations (\ref{field}) diverge more slowly
than the vacuum terms, $t^{-2(1+\delta )}$, as $t\rightarrow 0.$  This is the
condition for the Kasner solution to be unperturbed by the metric effects of
the fluid motions. We have to leading order that

\begin{eqnarray}
T_{0}^{0} \sim \rho u_{0}^{2}\sim t^{-1-2\delta -p_{3}} \; &; \qquad &
T_{1}^{1} \sim \rho \sim t^{-\frac{\gamma}{(2-\gamma)} (1+2 \delta-p_3)}
\label{t2} \\
T_{2}^{2} \sim \rho u_{2}u^{2}\sim t^{-2p_{2}\ -(1+2\delta -p_{3})}
\; &; \qquad
& T_{3}^{3} \sim \rho u_{3}u^{3}\sim t^{-1-2\delta -p_{3}}.  \label{t4}
\end{eqnarray}

\bigskip In the simple case of $\gamma =4/3$ where the 4-velocity of the
fluid is comoving ($u_{i}=\delta _{i}^{0}$) we require only that $\rho \sim
t^{-2-4\delta +2p_{3}}$ diverges slower than $t^{-2(1+\delta )}$ and, this
requires only that $p_{3}>\delta $, which is always satisfied. For general $%
\gamma $ the worst divergence created by relativistic motions is in $\rho
u_{3}u^{3}\sim t^{-1-2\delta -p_{3}}$ and this is always slower than $%
t^{-2(1+\delta )}$ since $p_{3}<1.$ But for stiffer equations of state we
have to careful that the assumption that the velocities are increasingly
relativistic $(u^{0}>>1)$ as $t\rightarrow 0$ continues to hold. This
requires $\gamma -1+2\delta <p_{3}$. If this inequality fails then $%
u^{\alpha }u_{\alpha }\rightarrow 0$ and $u^{0}\sim 1$ as $t\rightarrow 0$
and we need to re-solve (\ref{mo1})-(\ref{mo2}) in the approximation where
the velocities $u_{\alpha }u^{\alpha }$ go to zero and $u^{0}\simeq 1$
because the ultra-stiff fluid makes the motions grind to a halt \cite{jb5}%
. In this case $\rho \sim t^{-\gamma (1+2\delta )}$ and $\ u_{\alpha }\sim
t^{(\gamma -1)(1+2\delta )}$. Since we have $u_{\alpha }u^{\alpha }\sim
(u_{3})^{2}t^{-2p_{3}}\sim t^{2(\gamma -1)(1+2\delta )-2p_{3}}$ this
behaviour occurs when

\begin{equation}
(\gamma -1)(1+2\delta )>p_{3}.  \label{a}
\end{equation}
Since $u_{\alpha }u^{\alpha }\rightarrow 0$ as $t\rightarrow 0$ we will
always have $\rho >>\rho u_{\alpha }u^{\beta }$ as $t\rightarrow 0$ and we
need only check that $\rho $ diverges more slowly than the vacuum terms $%
O(t^{-2(1+\delta) })$ in order to check that the Kasner solution for the
metric is unperturbed by the fluid. This requires $(\gamma -2)(2\delta +1)<0$
and since $\gamma <2$ we require $2\delta +1>0$ for the vacuum term to
dominate.  This is just the condition for the universe volume to be
expanding as $t$ increases, (\ref{con3}), and always holds.

Hence, the Kasner solutions we have found will provide a good
description of the general spatially homogeneous perfect-fluid solutions of
the $R^{1+\delta }$ gravity theories in the vicinity of an initial
cosmological solution. They may also provide a useful approximation to the
time dependence of a general inhomogeneous cosmological solution to these
theories.

In summary, we have found new anisotropic vacuum solutions for the
scale-invariant gravity theories which generalise Einstein's general
relativity to a Lagrangian $R^{1+\delta }$. These solutions are expanding
universes of Kasner form with an initial spacetime singularity at $t=0$ and
exist if $-1/2<\delta <1/4$ but have different Kasner-index relations to the
classic Kasner solution of general relativity when $\delta \neq 0$. These
solutions are unperturbed by the introduction of non-comoving fluid matter
motions if $p<\rho $ on approach to the singularity for this range of $%
\delta ,$ and do not exhibit an infinite sequence of chaotic Mixmaster
oscillations when $\delta >0$. They should provide a simple new testing
ground for quantum cosmological processes and late-time behaviour in
theories `close' to Einstein's general relativity.

\ack

TC is supported by the PPARC.

\end{document}